# Plasma Heating Induced by Tadpole-Like Downflows in the Flaring Solar Corona

Tanmoy Samanta[1,2,9,*], Hui Tian[3,4,*], Bin Chen[5], Katharine K. Reeves[6], Mark C. M. Cheung[7], Angelos Vourlidas[2], Dipankar Banerjee[8]

[1]Department of Physics and Astronomy, George Mason University, Fairfax, Virginia 22030, USA.

[2]Johns Hopkins University Applied Physics Laboratory, Laurel, Maryland 20723, USA.

[3]School of Earth and Space Sciences, Peking University, Beijing 100871, People's Republic of China.

[4]Key Laboratory of Solar Activity, National Astronomical Observatories, Chinese Academy of Sciences, Beijing 100012, People's Republic of China.

[5]New Jersey Institute of Technology, Newark, New Jersey 07102, USA.

[6]Harvard-Smithsonian Center for Astrophysics, Cambridge, Massachusetts 02138, USA.

[7]Lockheed Martin Solar and Astrophysics Laboratory, Palo Alto, California 94304, USA.

[8]Aryabhatta Research Institute of Observational Sciences, Beluwakhan, Uttarakhand 263001, India.

[9]Part of this work was done during the postdoc appointment of T.S. at Peking University, Beijing, People's Republic of China

*Correspondence: tsamanta@gmu.edu (T.S.); huitian@pku.edu.cn (H.T.)

## Summary

**As one of the most spectacular energy release events in the solar system, solar flares are generally powered by magnetic reconnection in the solar corona. As a result of the re-arrangement of magnetic field topology after the reconnection process, a series of new loop-like magnetic structures are often formed and are known as flare loops. A hot diffuse region, consisting of around 5–10 MK plasma, is also observed above the loops and is called a supra-arcade fan. Often, dark, tadpole-like structures are seen to descend through the bright supra-arcade fans. It remains unclear what role these so-called supra-arcade downflows (SADs) play in heating the flaring coronal plasma. Here we show a unique flare observation, where many SADs collide with the flare loops and strongly heat the loops to a temperature of 10–20 MK. Several of these interactions generate clear signatures of quasi-periodic enhancement in the full-Sun-integrated soft X-ray emission, providing an alternative interpretation for quasi-periodic pulsations that are commonly observed during solar and stellar flares.**

## Introduction

Solar flares are characterized by a sudden enhancement of the electromagnetic radiation in a broad range of wavelengths on the Sun.[1] They are believed to be one of the dominant sources of severe disturbances in the space environments of the Earth and other planets in the solar system. Flares are generally believed to result from magnetic reconnection, a physical process referring to the re-arrangement of the magnetic field topology in a plasma. Through magnetic reconnection, the free magnetic energy is released and converted into kinetic and thermal energies.[1] As a result of the re-arrangement of the magnetic field topology, a series of new loop-like magnetic structures called flare loops are often formed below the reconnection site. A hot diffuse region, consisting of plasma with a temperature of around 5–10 MK, is also observed above the loops and is called a supra-arcade fan.[2–4]

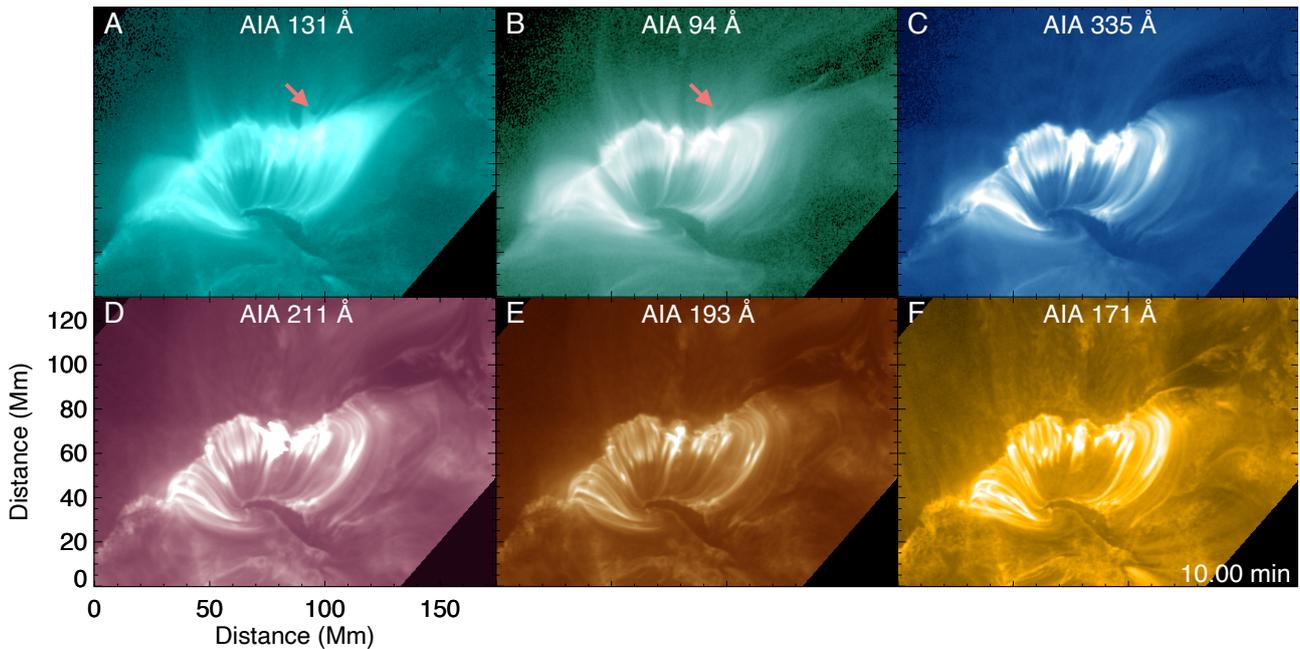

**Figure 1. Flare Loops Observed in Different AIA Filters.** The dark void-like SADs are best seen in AIA 131 Å and 94 Å images (A and B). One of the SADs is indicated by the red arrows in (A) and (B). The starting time of our observation is 07:40 UT on 2013 April 11. An animated version of (A) and (B) is available as a supplementary Video (Video S1).

Dark tadpole-like structures are often observed to descend through the bright supra-arcade fans.[5–7] Although their nature is debated,[5,6,8–10] these so-called supra-arcade downflows (SADs) are generally believed to be related to the downward movement of either reconnection outflows or rapid contraction of newly formed magnetic loops during the intermittent and turbulent magnetic reconnection process in flares. SADs are possibly hotter and less dense than the surrounding plasma.[8,11–14] They likely play an important role in shaping the plasma dynamics in supra-arcade fans; e.g., being involved in the generation of vortex shedding.[15] It is reported that SADs sometimes cause heating in the supra-arcade fans due to the compression of plasma.[16,17] However, these previously observed SADs generally do not reach the flare loops. Thus, it remains unclear what role SADs may play in heating the flare loops and the plasma in the immediate surrounding area.

In the past, SADs have been mostly observed above flare loops in the off-limb corona[18,19] when the reconnection current sheets face toward the observers. They were rarely observed on the solar disk,[20] possibly due to unfavorable geometry or reduced contrast with respect to the background emission.

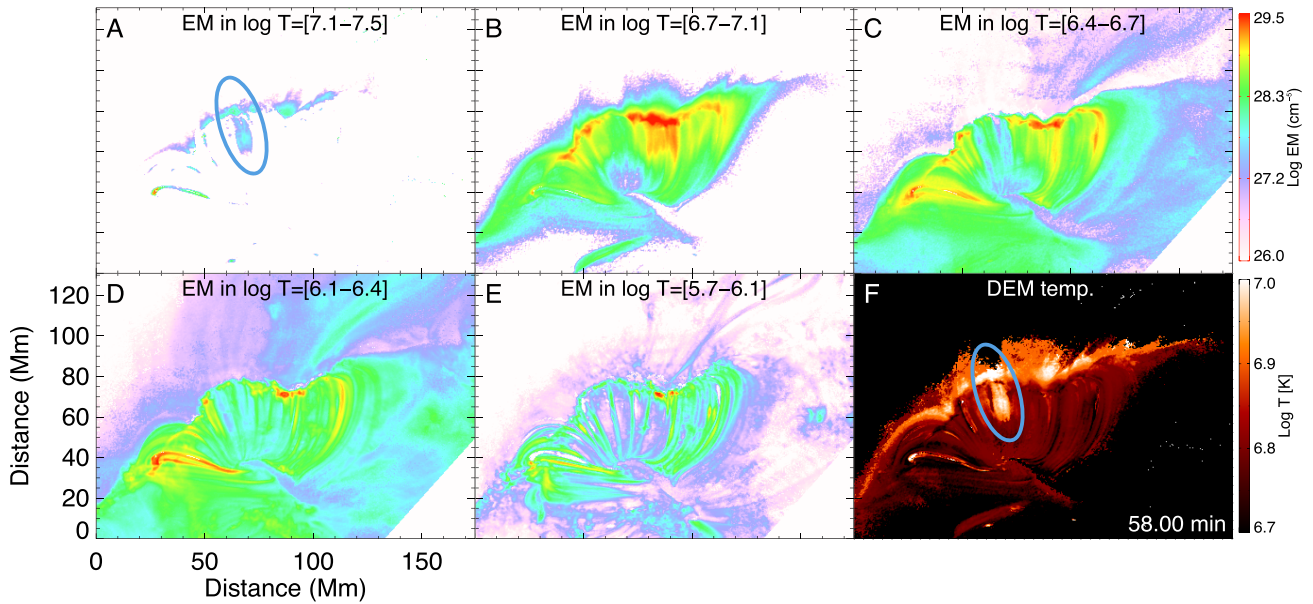

**Figure 2. EM and Temperature Maps.** (A–E) EM maps in different temperature bins. (F) Map of the DEM-weighted mean temperature. The blue ellipse highlights the location of hot plasma generated from the collision of an SAD with the loops. An animated version of this figure is available as a supplementary Video (Video S2).

# Results and Discussion

On April 11, 2013, we observed an M6.5-class flare in National Oceanic and Atmospheric Administration (NOAA) active region (AR) 11,719 (Figure 1, Video S1) on the solar disk by the Atmospheric Imaging Assembly (AIA[21]) on board the Solar Dynamics Observatory (SDO). Figure 1 shows images of the flare region at 7:50 UT as observed in different AIA filters. Most of these filters capture emission from plasma with a temperature of ~0.6–3 million degrees, whereas the 131 Å and 94 Å filters are sensitive to hotter plasma with temperatures of ~10 MK and 6 MK, respectively. Newly formed bright post-flare loops were observed in all of these six AIA filters. We also witnessed numerous downward-propagating SADs in the supra-arcade fan above the flare loops during the decay phase of the flare (mostly in the AIA 131 Å image sequence, some also seen in AIA 94 Å, Video S1). The favorable geometry and viewing angle of this event allowed us to observe the collision of many SADs with the flare loops and investigate the effect of collision in the local plasma heating.

## Plasma heating due to collision of the SADs with flare loops

To investigate possible plasma heating caused by the interactions, we focused on emission from the AIA hot channels (AIA 131 Å and AIA 94 Å) and we also performed a differential emission measure (DEM[22]; Figure 2, Video S2) analysis using images taken in the six Fe-dominated AIA filters (discussed later). Past observations showed that SADs mostly disappeared in supra-arcade fans before reaching flare loops. In contrast to previous cases, our on-disk observation shows that many SADs reach the apexes of flare loops and collide with them (Figure 3, Videos S1 and S3). When colliding with the loops, these SADs often reveal a sudden discontinuity in the downward propagation speed, from ~250 km/s in the supra-arcade fan region to ~15 km/s after the collision. Upward motion of hot plasma with a speed of ~15 km/s can also be identified from some collision sites. This rebound motion may be driven by the tension force associated with the distorted magnetic field lines at the loop tops or could be associated with the backflow resulting from the impact of the reconnection downflow on the strong magnetic field region, as predicted by numerical simulations.[23]

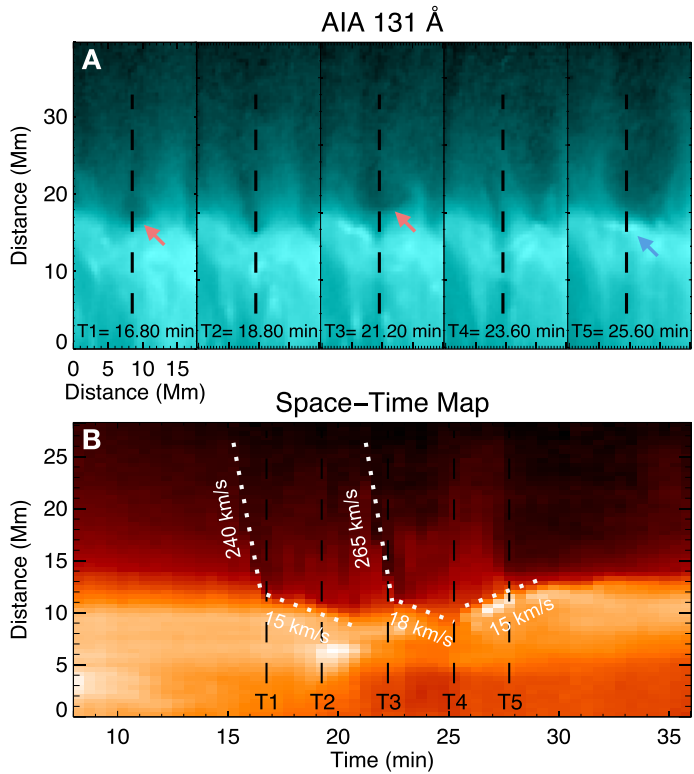

**Figure 3. Collision between SADs and Flare Loops.** (A) Image sequence of 131 Å showing the arrival of two SADs (marked by the two red arrows). The blue arrow indicates plasma rebounding from the interaction site. (B) A space-time (ST) diagram for the dashed line shown in (A). The slanted dashed lines mark the trajectories of the SADs and the rebound material, the slopes are used to estimate the propagation speeds. An animated version is available as a supplementary Video (Video S3).

When these SADs collide with the loops, they often create propagating intensity fluctuations along the flare loops. These fluctuations reach the foot-points of the loops and lead to brightenings at the footpoint areas (Video S4). We cannot derive the propagation speed in all cases due to the limited cadence. In a few cases, we noticed a time delay of ~24–60 s between the collision and footpoint brightening, which implies a propagation speed of 500–1,500 km/s.

Signatures of plasma heating have been found after the interactions, and Figure 4 shows one example (also see Video S5). We found that, when an SAD collides with the loops, the emission from the vicinity of the interaction site as well as the interacted loops suddenly decreases in all the six Fe-dominated AIA channels. A few minutes later, the intensity of AIA 131 Å starts to increase, followed by an increase of the 94 Å intensity and then the 335 Å intensity. The AIA 131 Å intensity peak is observed around 10–15 min after the collision. The intensity of AIA 94 Å peaks around 5–10 min after the peak of AIA 131 Å intensity. Similar behavior is observed both in the interacted loops (Figure 4J) and just above these loops (Figure 4I). The DEM-weighted mean temperature (Figures 2F, 4D, and 4H) shows a strong increase both at the interacted loops and slightly above the apexes of the loops around 10–15 min after the collision, and the temperature increase lasts for about 10–15 min. The apexes of the loops, where most of the energy is released from the collision, appear to be hotter (~15–20 MK; also see Figure 5) compared with the temperature of the heated loops (~10 MK). Some earlier observations[24] also suggested the presence of such hot (~10–20 MK) plasma at the loop tops during the flare decay phase, accompanied by impulsive microwave and X-ray bursts.

The Reuven Ramaty High Energy Solar Spectroscopic Imager (RHESSI[25]) observation of our flare revealed the presence of enhanced X-ray emission at the interaction locations and along the interacted loops (Figure 5). The X-ray spectra observed by RHESSI appear to be thermal, and a spectral fitting reveals a temperature of ~11–13 MK at the time of collision. Note that the integration time of RHESSI data is much longer (~4 min) and the spatial resolution is ~3 times lower compared with those of the AIA data. These differences may cause an underestimation in the temperature derived from the RHESSI data compared with the measurement from the high-cadence and high-resolution AIA data.

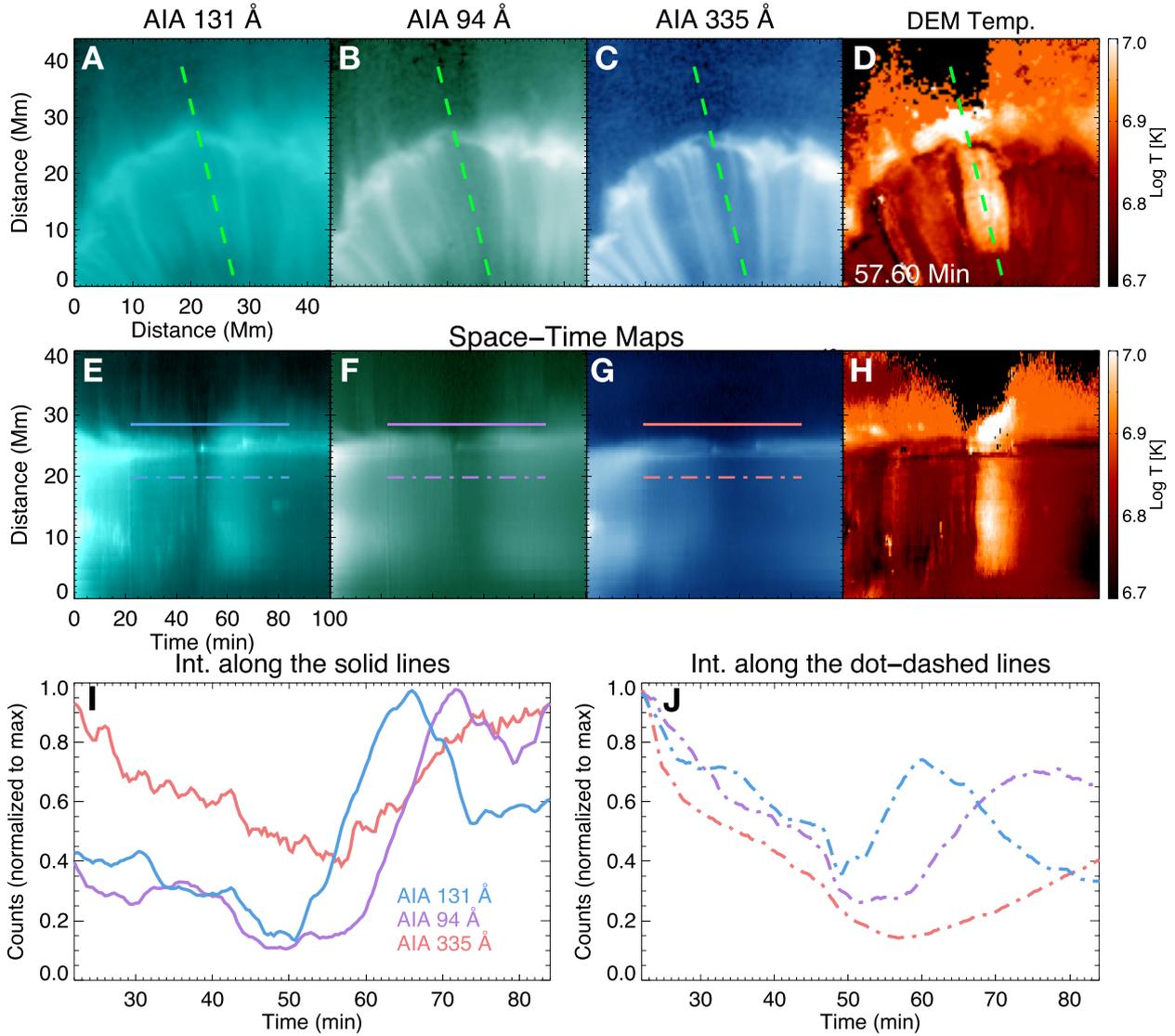

**Figure 4. An Example of Plasma Heating due to Interaction between SADs and Flare Loops.** (A–D) AIA 131 Å, 94 Å, 335 Å images and the map of DEM-weighted mean temperature after the collision of an SAD with the loops. (E–H) Temporal evolution along the green line (ST maps) in (A)–(D). The intensity variations along the solid and dot-dashed lines are shown in (I) and (J), respectively. An animated version is available as a supplementary Video (Video S5).

The strong velocity discontinuity of SADs might indicate the presence of shocks at the loop tops (e.g., termination shocks[23,26–29]). In this case, sudden dissipation of the kinetic energy at the shock locations could be responsible for the strong localized heating around the loop tops. It is also possible that the significant heating results from a strong compression during the collisions.[16,17] Thermal conduction could then rapidly spread the heating to the rest of the loops. If the plasma is heated to a temperature of >20 MK, the emission from all Fe-dominated AIA channels should decrease because these channels are generally insensitive to such a high temperature. This may be the reason for the sudden decrease in the intensities of all the Fe-dominated AIA filters just after the collisions (Figure 4). We realize that the 193 Å filter has a response around 18 MK, which is much weaker than the filter's response around 1.5 MK. If the ~18 MK plasma is much less than the 1.5 MK background coronal plasma in the line of sight, it will be very difficult to see an enhancement in the AIA 193 Å emission. Alternatively, the released energy could be immediately transported to electrons by thermal conduction (typical timescale << 1 s), which also leads to a non-equilibrium ionization condition[30] in the local plasma. During this period, the emission in all the Fe-dominated AIA filters may decrease due to the reduced efficiency of collisional excitation. Later, the ions could be heated through collisions with the slowly cooling hot electrons. Therefore, the spectral line emission from hot ions starts to increase after thermal relaxation between electrons and ions. After the AIA 131 Å intensity reaches its peak, the plasma starts to cool down, leading to intensity peaks in 94 Å and then in 335 Å. In addition, non-thermal particles may also play a role in the heating, particularly if the SADs are due to downward propagating plasmoids from the reconnection site.

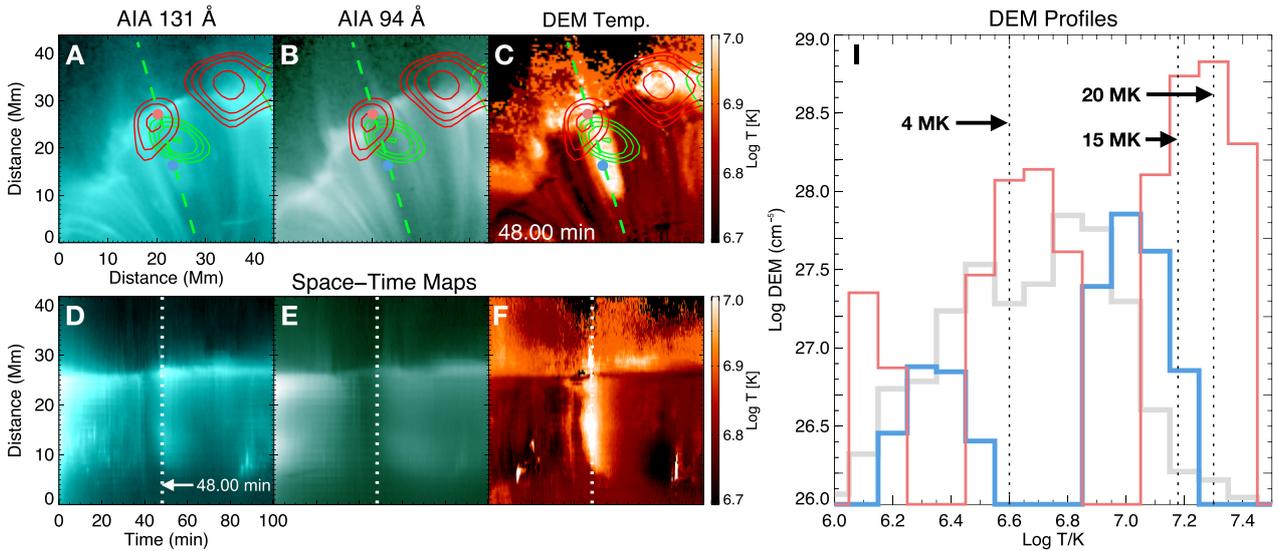

**Figure 5. Another Example of Plasma Heating due to Interaction between SADs and Flare Loops.** (A–C) AIA 131 Å and 94 Å images, and map of DEM-weighted mean temperature after the collision of an SAD with the loops. The red and green contours represent the 10–20 keV and 6–10 keV X-ray sources observed with RHESSI, respectively. (D–F) Temporal evolution along the green line (ST maps) in (A)–(C). The white dashed line indicates the time instant of the images shown in (A)–(C). (I) DEM profiles. The gray curve is the DEM curve averaged over the whole region of (A). The blue and red curves represent the DEM curves at the locations marked by the blue and red dots in (A)–(C), respectively. The black dotted lines indicate temperatures of 4 MK, 15 MK, and 20 MK.

## Quasi-periodic pulsations caused by SADs

We further analyzed the time variations of the AIA 131 Å intensity and DEM temperature (averaged over the whole region of Figure 6A) along with the soft X-ray fluxes measured by the Geostationary Operational Environmental Satellites (GOES) during the post-flare phase. These light curves all show clear quasi-periodic variations (Figures 6B and 6C). We performed the wavelet analysis[31] for the time series of AIA 131 Å intensity and DEM temperature, and found a period of around 10 min for both of them (Figure S1). We also found that the GOES fluxes exhibit a similar behavior. Several peaks in the GOES data appear to occur at the same times as the AIA 131 Å intensity and DEM temperature peaks.

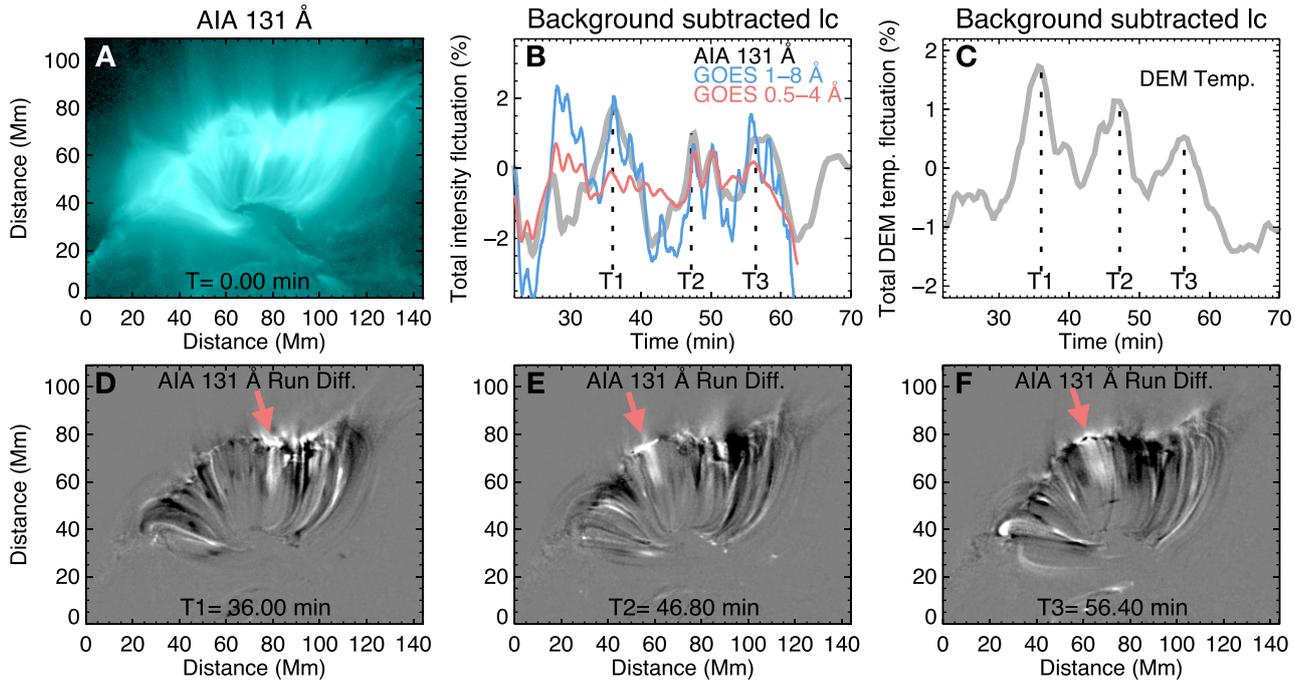

**Figure 6: QPP Caused by Collisions of SADs with Flare Loops.** (A) An image of AIA 131 Å taken at 07:40 UT. (B) Variations of the AIA 131 Å intensity (integrated over the whole region of panel A) and GOES X-ray fluxes. (C) Variations of the DEM-weighted mean temperature averaged over the whole region of panel A. All light curves are subtracted and normalized by a 20-min smoothed background. (D-F) Running difference images of AIA 131 Å at the time instances T1, T2 and T3 (marked in panels B and C), respectively. Three red arrows indicate the locations of strong intensity enhancement due to collisions of SAD with the loops. An animated version is available as an online supplementary movie (Movie S6).

Quasi-periodic intensity variations have been frequently observed during many solar and stellar flares,[32–37] and they are often referred to as quasi-periodic pulsations (QPPs). These QPPs are often explained as being caused by magnetohydrodynamic (MHD) oscillations of the loops[23,38,39] or associated with repetitive/intermittent magnetic reconnection during flares.[40,41] Our detailed investigation shows that several peaks of the QPPs are clearly related to the interaction of SADs with the flare loops and the resultant heating. Three such peaks are marked by the three dashed lines (T1–T3) in the curves shown in Figures 6B and 6C. The three corresponding heating signatures could be identified as local intensity enhancements in the AIA 131 Å images (Figures 6D–6F, Video S6). Hence, our observations have revealed an alternative scenario for the generation of QPPs in the decay phase of some flares. Recent numerical simulations found that plasmoid collisions with flare loops could lead to turbulent heating and QPPs above the loop tops.[42] Our observations appear to be consistent with this scenario, although it is unclear whether the observed SADs are plasmoids in the reconnection current sheet.

# Conclusion

Mysterious dark tadpole-like downflows (SADs) are often observed at flare regions in the solar corona. Most of the previously reported SADs were observed above the solar limb. These observations showed that these Sun-ward-propagating SADs mostly disappear in the diffuse plasma region of flares before reaching the newly formed flare loops. Our observation of a flare that occurs on the solar disk reveals that many SADs not only reach the flare loops but also collide with these loops and heat the plasma to a temperature of about 10–20 MK. Since different SADs collide with the loops at different occasions, the plasma heating and resultant Extreme-UV (EUV)/X-ray emission enhancement occurs quasi-periodically. Hence, our observation reveals that SADs play an important role in heating the coronal plasma in flare regions and also provides an alternative interpretation for the highly debated nature of frequently observed QPPs in solar and stellar flares.

# MATERIAL AND METHOD

## Observations and data reduction

We used data from the AIA instrument on board the SDO spacecraft, the GOES, and the RHESSI satellites.

The AIA instrument provides full-disk solar images of the Sun. In this study, we used images taken in the six Fe-dominated EUV passbands (94 Å, 131 Å, 171 Å, 193 Å, 211 Å, 335 Å) during the period of 07:40 UT to 09:20 UT on 2013 April 11. These passbands sample plasma with different temperatures in the solar atmosphere. AIA images are calibrated and co-aligned onto the same plate-scale using the aia_prep.pro routine in SolarSoftWare (SSW). The pixel size of these AIA images is 0.6". Although each filter took images at a 12-s cadence, we used data with a 24-s cadence in our study since we found that many of the alternate frames are saturated at different locations.

The GOES satellite measured soft X-ray fluxes of the full Sun in two wavelength bands (0.5–4 Å and 1–8 Å) during the same time period. We also used the reconstructed RHESSI X-ray images in two energy bands (6–10 keV and 10–20 keV). The pixel size of the RHESSI images is 2.0 inches and the integration time selected for imaging is 2–4 min depending on the count rate. The GOES X-ray fluxes were obtained using the SSW software GOES Workbench. The RHESSI images were processed through SSW using subcollimators 3, 4, 5, 6, 7, 8. RHESSI spectroscopy was performed using subcollimator 3 with an integration time of 48 s. The X-ray spectral fit was per- formed using the SSW OSPEX package by adopting an isothermal model within an energy range of 6 keV–15 keV (above which the background becomes significant). The spectral fit of the X-ray source returns the plasma temperature T and a volume emission measure EM_V.

## DEM analysis

We applied a widely used inversion method[22] to AIA images taken in the six Fe-dominated passbands to compute the DEM curve at each spatial pixel and time step. DEM represents the amount of plasma as a function of temperature. The inversion method uses the temperature response functions of the six AIA channels and the measured count rates (in DN/s) in different channels. We have also applied a modified version of this method,[43] and obtained similar results. Hence, here we only present results from the former version.[22]

The DEM curves in the flare region generally show a double-peak distribution,[44,45] consisting of a cold component peaking around log T/K ~ 6.2 and a hot component that peaks around log T/K ~ 7.0. The cold component is mainly from the emission of the regular coronal plasma (background emission), and the hot component arises from the heating due to magnetic energy release from the reconnection. It is proposed that the mean weighted temperature of the hot DEM component is a better representation of the temperature of the flaring plasma.[44,45] The DEM-weighted mean temperature is conventionally defined as follows,

$$<T>_h = \frac{\sum DEM(T) \times T \Delta T}{\sum DEM(T) \times \Delta T} \qquad \text{(Equation 1)}$$

The integration was performed in the temperature range of 4–32 MK for investigating the dynamics of the hot plasma.[44] As an example, Figure 2 shows images of the emission measure (EM) in different temperature bins as well as the calculated $<T>_h$ (also see Video S2). Here, EM refers to the integral of DEM(T) over a finite range of temperature.

# References

1. Benz, A.O. (2008). Flare observations. Living Rev. Sol. Phys. 5, 1.
2. McKenzie, D.E., and Hudson, H.S. (1999). X-ray observations of motions and structure above a solar flare arcade. Astrophys. J. Lett. 519, L93–L96.
3. Verwichte, E., Nakariakov, V.M., and Cooper, F.C. (2005). Transverse waves in a post-flare supra-arcade. Astron. Astrophys. 430, L65–L68.
4. Hanneman, W.J., and Reeves, K.K. (2014). Thermal structure of current sheets and supra-arcade downflows in the solar corona. Astrophys. J. 786, 95.
5. McKenzie, D.E. (2000). Supra-arcade downflows in long-duration solar flare events. Sol. Phys. 195, 381–399.
6. Savage, S.L., and McKenzie, D.E. (2011). Quantitative examination of a large sample of supra-arcade downflows in eruptive solar flares. Astrophys. J. 730, 98.
7. Li, L.P., Zhang, J., Su, J.T., et al. (2016). Oscillation of current sheets in the wake of a flux rope eruption observed by the solar dynamics observatory. Astrophys. J. Lett. 829, L33.
8. Savage, S.L., McKenzie, D.E., and Reeves, K.K. (2012). Re-interpretation of supra-arcade down-flows in solar flares. Astrophys. J. Lett. 747, L40.
9. Cassak, P.A., Drake2, J.F., and Gosling, J.T. (2013). On the cause of supra-arcade downflows in solar flares. Astrophys. J. Lett. 775, L14.
10. Guo, L.-J., Huang, Y.-M., Bhattacharjee, A., et al. (2014). Rayleigh-Taylor type instabilities in the reconnection exhaust jet as a mechanism for supra-arcade downflows in the sun. Astrophys. J. Lett. 796, L29.
11. Innes, D.E., McKenzie, D.E., and Wang, T. (2003). SUMER spectral observations of post-flare supra-arcade inflows. Sol. Phys. 217, 247–265.
12. Maglione, L.S., Schneiter, E.M., Costa, A., et al. (2011). Simulation of dark lanes in post-flare supra-arcades. III. A 2D simulation. Astron. Astrophys. 527, L5.
13. Cécere, M., Schneiter, M., Costa, A., et al. (2012). Simulation of descending multiple supra-arcade reconnection outflows in solar flares. Astrophys. J. 759, 79.
14. Cheng, X., Li, Y., Wan, L.F., et al. (2018). Observations of turbulent magnetic reconnection within a solar current sheet. Astrophys. J. 866, 64.
15. Samanta, T., Tian, H., and Nakariakov, V.M. (2019). Evidence for vortex shedding in the sun's hot corona. Phys. Rev. Lett. 123, 035102.
16. Reeves, K.K., Török, T., Mikic´, Z., et al. (2019). Exploring plasma heating in the current sheet region in a three-dimensional coronal mass ejection simulation. Astrophys. J. 887, 103.
17. Xue, J., Su, Y., Li, H., et al. (2020). Thermodynamical evolution of supra-arcade down-flows. Astrophys. J. 898, 88.
18. Savage, S.L., McKenzie, D.E., Reeves, K.K., et al. (2010). Reconnection outflows and current sheet observed with Hinode/XRT in the 2008 April 9 "Cartwheel CME" flare. Astrophys. J. 722, 329–342.
19. Warren, H.P., O'Brien, C.M., and Sheeley, N.R., Jr. (2011). Observations of reconnecting flare loops with the atmospheric imaging assembly. Astrophys. J. 742, 92.
20. Innes, D.E., Guo, L., Bhattacharjee, A., et al. (2014). Observations of supra-arcade fans: instabilities at the head of reconnection jets. Astrophys. J. 796, 27.
21. Lemen, J.R., Title, A.M., Akin, D.J., et al. (2012). The atmospheric imaging assembly (AIA) on the Solar Dynamics Observatory (SDO). Sol. Phys. 275, 17–40.

22. Cheung, M.C.M., Boerner1, P., Schrijver, C.J., et al. (2015). Thermal diagnostics with the atmospheric imaging assembly on board the solar dynamics observatory: a validated method for differential emission measure. Inversions. Astrophys. J. 807, 143.
23. Takasao, S., and Shibata, K. (2016). Above-the-loop-top oscillation and quasi-periodic coronal wave generation in solar flares. Astrophys. J. 823, 150.
24. Yu, S., Chen, B., Reeves, K.K., et al. (2020). Magnetic reconnection during the post- impulsive phase of a long-duration solar flare: bidirectional outflows as a cause of microwave and X-ray bursts. Astrophys. J. 900, 17.
25. Lin, R.P., Dennis, B.R., Hurford, G.J., et al. (2002). The Reuven Ramaty High-Energy Solar Spectroscopic Imager (RHESSI). Sol. Phys. 210, 3–32.
26. Chen, B., Bastian, T.S., Shen, C., et al. (2015). Particle acceleration by a solar flare termination shock. Science 350, 1238–1242.
27. Polito, V., Galan, G., Reeves, K.K., et al. (2018). Possible signatures of a termination shock in the 2014 March 29 X-class flare observed by IRIS. Astrophys. J. 865, 161.
28. Shen, C., Kong, X., Guo, F., et al. (2018). The dynamical behavior of reconnection-driven termination shocks in solar flares: magnetohydrodynamic simulations. Astrophys. J. 869, 116.
29. Cai, Q., Shen, C., Raymond, J.C., et al. (2019). Investigations of a supra-arcade fan and termination shock above the top of the flare-loop system of the 2017 September 10 event. Mon. Not. Roy. Astron. Soc. 489, 3183–3199.
30. Shen, C., Revees, K.K., Raymond, J.C., et al. (2013). Non-equilibrium ionization modeling of the current sheet in a simulated solar eruption. Astrophys. J. 773, 110.
31. Torrence, C., and Compo, G.P. (1998). A practical guide to wavelet analysis. Bull. Am. Meteorol. Soc. 79, 61.
32. Mathioudakis, M., Seiradakis, J.H., Williams, D.R., et al. (2003). White-light oscillations during a flare on II Peg. Astron. Astrophys. 403, 1101–1104.
33. Zimovets, I.V., Kuznetsov, S.A., and Struminsky, A.B. (2013). Fine structure of the sources of quasi-periodic pulsations in "single-loop" solar flares. Astron. Lett. 39, 267–278.
34. Van Doorsselaere, T., Kupriyanova, E.G., and Yuan, D. (2016). Quasi-periodic pulsations in solar and stellar flares: an overview of recent results (invited review). Sol. Phys. 291, 3143–3164.
35. Pugh, C.E., Armstrong, D.J., Nakariakov, V.M., et al. (2016). Statistical properties of quasi-periodic pulsations in white-light flares observed with Kepler. Mon. Not. Roy. Astron. Soc. 459, 3659–3676.
36. McLaughlin, J.A., Nakariakov, V.M., Dominique, M., et al. (2018). Modelling quasi-periodic pulsations in solar and stellar flares. Space Sci. Rev. 214, 45.
37. Hayes, L., Gallagher, P.T., Denni, B.R., et al. (2019). Persistent quasi-periodic pulsations during a large X-class solar flare. Astrophys. J. 875, 33.
38. Nakariakov, V.M., and Melnikov, V.F. (2009). Quasi-periodic pulsations in solar flares. Space Sci. Rev. 149, 119–151.
39. Tian, H., Young, P.R., Reeves, K.K., et al. (2016). Global sausage oscillation of solar flare loops detected by the interface region imaging spectrograph. Astrophys. J. Lett. 823, L16.
40. Li, D., and Zhang, Q.-M. (2017). Quasi-periodic pulsations with multiple periods in hard X-ray emission. Mon. Not. Roy. Astron. Soc. 471, L6–L10.
41. Yuan, D., Feng, S., Li, D., et al. (2019). A compact source for quasi-periodic pulsation in an M-class solar flare. Astrophys. J. Lett. 886, L25.
42. Ye, J., Cai, Q., Shen, C., et al. (2020). The role of turbulence for heating plasmas in eruptive solar flares. Astrophys. J. 897, 64.
43. Su, Y., Veronig, A.M., Hannah, I.G., et al. (2018). Determination of differential emission measure from solar extreme ultraviolet images. Astrophys. J. Lett. 856, L17.
44. Gou, T., Liu, R., and Wang, Y. (2015). Do all candle-flame-shaped flares have the same temperature distribution? Sol. Phys. 290, 2211–2230.
45. Chen, X., Liu, R., Deng, N., et al. (2017). Thermodynamics of supra-arcade downflows in solar flares. Astron. Astrophys. 606, A84.

## Acknowledgments

The authors thank the SDO, GOES and RHESSI teams for providing the data, and Shinsuke Takasao for helpful discussion. This work was supported by NSFC grants 11825301 and 11790304, Strategic Priority Research Program of CAS (grant XDA17040507), NASA LWS grant 80NSSC19K0069, NSF grants AST-1735405 and AGS-1723436 to NJIT, NASA grant 80NSSC18K0732 and NASA's SDO/AIA contract (NNG04EA00C) to the Lockheed Martin Solar and Astrophysics Laboratory. AIA is an instrument onboard the Solar Dynamics Observatory, a mission for NASA's Living With a Star program.

## Data availability

The AIA and HMI data are available at the Joint Science Operations Center (http://jsoc.stanford.edu/). The GOES and RHESSI data can be downloaded and processed through SSW (https://hesperia.gsfc.nasa.gov/rhessi3/).

## Declaration of interests

The authors declare no competing financial interests.

## Supplementary Movies

Supplemental information can be found online at https://doi.org/10.1016/j.xinn.2021.100083

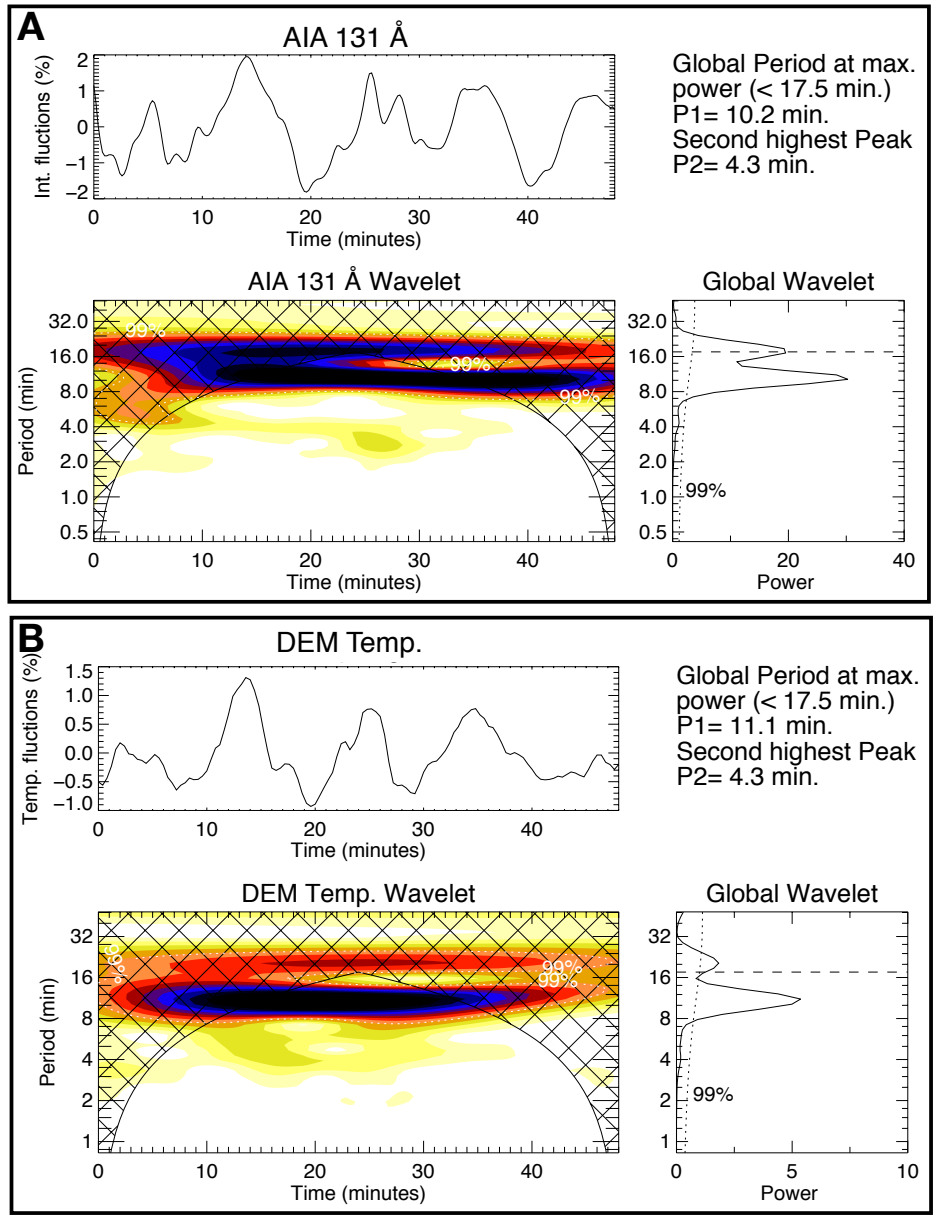

**Figure S1: Results of wavelet analysis for the time series of the AIA 131 Å intensity and DEM temperature averaged over the whole region of Figure 6A.** (A) Top panel shows the normalized intensity variation of AIA 131 Å. The bottom left panel shows the wavelet power spectrum. The cross-hatched region above the wavelet power spectrum highlights the cone of influence (COI). The location of power above 99.99 % significance level is represented by the region overplotted with dotted white lines. The bottom right panel shows the global wavelet power. The longest measurable period is 11.1 minutes (due to the COI), which is indicated by a horizontal dashed line. The dotted line shows the significance level of 99.99 %. The significant periods that are measured from the global wavelet power are printed at the top of the global wavelet power plot. (B) The panels are the same as in (A) but for the DEM temperature.